\documentclass[preprint2]{aastex}
\def\msol{M_\odot}

\def\simgr{\,\hbox{\hbox{$ > $}\kern -0.8em \lower 1.0ex\hbox{$\sim$}}\,}
\def\simle{\,\hbox{\hbox{$ < $}\kern -0.8em \lower 1.0ex\hbox{$\sim$}}\,}
\def\msolyr{M_\odot {\rm yr^{-1}}}

\def\te{T_{\rm eff}}
\def\msolyr{{\rm M}_\odot {\rm yr^{-1}}}
\lefthead{THORSTENSEN ET AL.}
\righthead{QZ Ser}
\begin{document}
\title{QZ Serpentis: A Dwarf Nova with a 2-Hour Orbital 
Period and an Anomalously Hot, Bright Secondary Star 
\footnote{Based in part on
observations obtained at the MDM Observatory, operated by
Dartmouth College, Columbia University, Ohio State University, and
the University of Michigan.}
}

\author{John R. Thorstensen, William H. Fenton}
\affil{Department of Physics and Astronomy\\
6127 Wilder Laboratory, Dartmouth College\\
Hanover, NH 03755-3528;\\
john.thorstensen@dartmouth.edu, w.h.fenton@dartmouth.edu}
\author{Joseph O. Patterson, Jonathan Kemp\altaffilmark{2}, Jules Halpern}
\affil{Department of Astronomy, Columbia University\\
538 West 120th Street, New York, NY 10027;\\
jop@astro.columbia.edu, j.kemp@jach.hawaii.edu,jules@astro.columbia.edu}
\author{Isabelle Baraffe}
\affil{\'Ecole Normale Sup\'erieure, 69364 Lyon Cedex 07, France;\\
ibaraffe@ens-lyon.fr}
\altaffiltext{2}{Also at Joint Astronomy Center, Hilo, Hawaii.}

\begin{abstract}
We present spectroscopy and time-series photometry of 
the dwarf nova QZ Ser.  The spectrum shows a rich absorption
line spectrum of type $K4 \pm 2$.  K-type secondary stars 
are generally seen in dwarf novae with orbital periods 
$P_{\rm orb} \sim 6$ h, but
in QZ Ser the absorption radial velocities show 
an obvious modulation (semi-amplitude $207(5)$ km s$^{-1}$) at
$P_{\rm orb} = 119.752(2)$ min, much shorter than typical for such a
relatively warm and prominent secondary spectrum.
The H$\alpha$ emission-line velocity is
modulated at the same period and roughly opposite phase.
Time-series photometry shows flickering superposed
on a modulation with two humps per orbit, consistent with
ellipsoidal variation of the secondary's light.  
QZ Ser is a second example of a relatively short-period
dwarf nova with a surprisingly warm secondary.  Model
calculations suggest that the secondary is strongly enhanced
in helium, and had already undergone significant nuclear
evolution when mass transfer began.  Several sodium absorption 
features in the secondary spectrum are unusually strong, which may
indicate that the present-day surface was the
site of CNO-cycle hydrogen burning in the past.
\end{abstract}

\section{Introduction}

Cataclysmic variable stars (CVs) are close binary systems
in which a low-mass secondary transfers mass onto a white dwarf;
\citet{warn} wrote an excellent monograph on CVs.

The Roche geometry tightly constrains the secondary star's mass
at a given orbital period $P_{\rm orb}$.  Short-period systems
have low-mass secondaries, so if the chemical composition  
is normal ($X \sim 0.7$), 
the secondary is a faint M dwarf or brown dwarf and contributes negligibly
to the visible-light spectrum (Fig.~4 of \citealt{patprecess01});
However, \citet{thor02} found a K4 $\pm 2$ secondary in the 
dwarf nova 1RXS J232953.9+062814 (hereafter RX 2329+06), 
which has $P_{\rm orb}$ = 64 min. 
They suggested that the secondary was somewhat
evolved at the start of mass transfer, with its core substantially
enhanced in helium.  In this scenario the portion of the secondary
remaining today corresponds to the core of the original star, and the 
enhanced helium greatly affects the mass-temperature relation.  

Here we present new observations of the dwarf nova QZ Ser, which 
appears to be a close relative of RX 2329+06. 
QZ Ser was discovered by Katsumi Haseda in 1998 and designated
as HadV04 in the discovery notification (vsnet-obs 18349)\footnote
{The vsnet mailing list archives are available at http://vsnet.kusastro.kyoto-u.ac.jp/vsnet/index.html}
, in which  
T. Kato suggested an identification with the ROSAT
WGA source 1WGAJ1556.9+2108 \citep{wga94}.  The outbursting object 
was discovered on small-scale patrol films, and its quiescent counterpart
was initially uncertain.  On 2001 June 28 we obtained a spectrum of the 
star nearest the position, and it proved to be an ordinary late-type star.
Returning to the field on 2002 January 20 we obtained
a spectrum of a somewhat fainter star 11 arcsec to the northwest.
This showed the distinctive broad Balmer emission characteristic
of dwarf novae at minimum light, together with absorption
features of a late-type star.   On 2002 Feb 3.23 UT, P. Schmeer 
detected an outburst of QZ Ser (vsnet-campaign 1281), and an examination 
of an outburst image by H. Yamaoka (vsnet-alert 7172) 
confirmed that the outbursting object was on the position of 
the emission-line object, cementing the identification.

The position of QZ Ser, derived from a fit to 62 USNO A2.0
stars \citep{mon96} on one of our 1.3m images (described below), is 
$\alpha_{\rm ICRS} = 15^{\rm h}\ 56^{\rm m}\ 54^{\rm s}.50,\  
\delta_{\rm ICRS} = +21^{\circ}\ 07'\ 19''.5\ (\pm 0.''3$ estimated
uncertainty).  Its position in the USNO A2.0 is not significantly
different, which sets an upper limit on its proper motion  
$\mu \simle 0''.03$ yr$^{-1}$.  
The {\it Living Edition} of the {\it Catalog and Atlas of Cataclysmic
Variables} \citep{downes2001} gives an up-to-date finding chart. 

\section{Observations and Analysis}

Our time-series photometry is from the 1.3 m Mcgraw-Hill telescope
at MDM Observatory on Kitt Peak, during 2002 February.  After
bias subtraction and flat fielding our $I$-band CCD pictures, we 
measured instrumental magnitudes using aperture photometry, and
differenced the program star against a nearby comparison object.

On 2002 January 21 we obtained UBVI exposures in photometric 
conditions with the 1.3 m telescope.  We measured
these with DAOPHOT and transformed to
standard UBVI magnitudes using observations of 
\citet{landolt92} standard stars.
Our BV observations agreed within a few hundredths of a
magnitude with photometry listed by Henden
\footnote{see ftp://ftp.nofs.navy.mil/pub/outgoing/aah/sequence/}.  QZ Ser
had $V = 17.69, B - V = 0.72, U - B = -0.32,$ and
$V - I_{\rm KC} = 1.22$, rather redder
than typical short-period dwarf novae at minimum light.

Table 1 gives a journal of our spectroscopy, all of which is
from the 2.4 m Hiltner telescope at MDM Observatory on Kitt Peak.  
The procedures, instrumental
setup and reduction were essentially as described 
in \citet{t98}.  Nearly all the exposures from 
the extensive 2002 February data set were 480 s, 
the total exposure in February being 38820 s.
Because of the season, the February data span only 
3.4 h of hour angle.  Four 480-s spectra were taken 
in 2002 June, after this paper was first submitted, mostly to  
improve the ephemeris.  

Fig.~1 shows the mean spectrum from February.  The mean
spectrum from January, before the outburst, appears
somewhat similar, but with a flux level
$\sim 30$ per cent lower in the $V$ band, a marginally
redder continuum, and much weaker He I emission, and
in June the spectrum resembled the one taken in January. 
Without simultaneous photometry it is difficult to be certain,
but in February the star may have been slightly brighter than its minimum,
perhaps because of the outburst earlier that month.
Table 1 details the emission lines in the February spectrum.
The lines appear 
double-peaked at the top, with separation $\sim 600$  km s$^{-1}$.  
Late-type absorption features are also present.

We measured absorption velocities by cross-correlating
our spectra against a rest-frame sum of late-type IAU velocity
standards, using {\it xcsao} \citep{kurtzmink98}.  The
6000 to 6500 \AA\ region gave the best results, with  
mean formal uncertainty 10 km s$^{-1}$.   We discarded
a few velocities with large formal errors,
leaving four velocities from 2002 January 
and 69 from 2002 February.  A least-squares period search 
\citep{tpst} showed a strong periodicity near
two hours.   Because the velocity uncertainties are
small compared to the amplitude, the period is determined
without ambiguity in the daily cycle count despite the relatively small 
hour-angle span of the data, and the cycle count between the 
January and February velocities is also secure. 
Table 3 gives sinusoidal fit parameters for the velocities, 
and Fig.~2 shows the velocities folded on the period.
H$\alpha$ emission-line velocities measured
using a convolution method sensitive to the line wings
\citep{sy} gave noisier velocities (also in Fig.~2), 
but yielded a consistent period. 
A fit to the emission-line velocities with the period fixed
at the more accurate absorption-line value gave parameters listed
in Table 3.  The phase of the emission line velocities lags 
the absorption lines
by 0.58(2) cycles (formal error), whereas a lag of exactly 
0.5 would be expected if the emission accurately traced the 
white-dwarf motion.  We conclude that, as often is the case,
the emission lines do not trace the white dwarf motion accurately.
The equal periods and roughly opposing phases of the emission and
absorption do confirm that they arise in the same system, and that the
absorption features are from the secondary and not some interloper.

Fig.~3 shows a greyscale representation of the February spectra
arranged as a function of phase \citep{tay98}.  The sharp 
absorption features and 
their modulation are obvious.  The HeI $\lambda$ 5876 line shows 
evidence of an $S$-wave, i.e., an emission component apparently 
arising from the hot spot where the accretion stream 
strikes the disk.  A phase-dependent absorption 
around $\lambda 5910$, with $\sim 20$  \AA\ FWHM 
is also clearly visible.  We measured the feature's equivalent
width in the original spectra, and while these measures
were noisy they showed a smooth and persistent modulation
with binary phase (Fig.~2, lower panel).  
A sine fit gave a mean EW of 3.8 \AA , with a
half-amplitude of 2.2 \AA .  The feature reaches maximum
strength around the time of maximum radial velocity of the
secondary.  It seems likely that this feature is due to sodium,
but we have been unable to explain it further.  Its diffuseness
suggests a velocity spread characteristic of 
the accretion structures rather than the 
stellar photosphere.  Although phase-dependent absorption calls
to mind the superficially similar phenomenon in SW Sex stars, the
morphology and phasing of this feature in the single-trailed 
spectrum is quite different from typical SW Sex absorption events
(see e.g. \citealt{tay98}).

To estimate the secondary's spectral type, we used
spectra (obtained with the same instrumental setup) of stars
classified by \citet{keenan89}.  Using two 
temperature-sensitive line ratios appropriate for our 
spectral resolution and coverage, we found
K4 $\pm$ 2 subclasses for QZ Ser's secondary.
Most of the absorption features appeared similar
to the spectral type standards, but the standards
do not show the broad $\lambda 5910$ feature noted above.
A pair of absorption lines at 
$\lambda\lambda 5683,5688$ also appears 
unusually strong, with equivalent widths of 
0.9 and 1.2 \AA\ respectively.  These are evidently
a Na I triplet at $\lambda\lambda$ 5682.63, 5688.19,
and 5688.21.

The strength of these lines prompted us to search for other
sodium features.  The NaD lines are unsuitable because of saturation
and confusion with $\lambda 5876$ (they may also be responsible
for the unusual $\lambda 5910$ feature), 
but a search of the NIST atomic line database
\footnote{http://physics.nist.gov/cgi-bin/AtData/main\_asd}
yields other features.  One, near at $\lambda 6154.225$, appears 
enhanced in equivalent width by a factor of about three compared to 
our K star standards; a neighboring sodium line at $\lambda 6160.747$ is 
unfortunately blended with a strong Ca I line at our resolution.  
In several other features (a blend of $\lambda\lambda 4978.5$ and 
4982.8, and one at $\lambda\lambda 4664.8$ and 4668.6) the enhancement
is not enough to be significant, though the features are clearly detected.  
All the sodium features considered here arise from the 3p level,
further corroborating the line identification.
These sodium features move with the secondary's spectrum, so 
we conclude that there is circumstantial evidence for an 
enhanced sodium abundance in the secondary's photosphere.   

Because the spectrum does not appear perfectly normal, 
we cannot put a strong constraint on the fraction of the light 
contributed by the secondary, but it is
clearly very substantial.  From systematically subtracting
the library spectra from the average spectrum, we 
estimate that the secondary contributes $70 \pm 20$ percent 
of the light in the 5500 -- 6500 \AA\ range.
On face value, our $V$ magnitude then implies $V = 18.1 \pm 0.4$ 
for the secondary alone.  Because the secondary's
fractional contribution was probably a little higher
in January (when the photometry was taken) than in
February (when the secondary contribution was measured),
our best estimate for the secondary's magnitude is
$V = 17.9 \pm 0.4$.

The lower panel of Fig.~2 shows averaged, mean-subtracted, 
differential $I$-band magnitudes as a function of orbital phase.  
The time averaging suppresses
considerable flickering present in the original
time series.  The light curve shows two humps per orbit,
with mean full amplitude 0.076(9) mag.  The minima 
appear to be asymmetric, differing by $\Delta A = 0.012(7)$ mag.
This is the waveform characteristic of tidally distorted secondaries within a
Roche lobe, as calculated for example by \citet{boch79} (hereafter BKS).

\section{Inferences}

{\it Distance.} The secondary spectrum and period
constrain the distance.  Assuming the secondary
fills its Roche lobe, we have 
$R_2/R_{\odot} = 0.234 f(q) P_{\rm hr}^{2/3} {(M_2 / M_{\odot})}^{1/3}$
\citep{beuermann98}, where $f(q)$ is very close to unity over
the range of interest.  
Evolutionary models suggest $M_2 = 0.125$ M$_{\odot}$ (see below), and
because this enters weakly in the present calculation we take this
as a guide, assuming an uncertainty of $\pm 0.025$ M$_{\odot}$,
which yields $R_2 = 0.185 \pm 0.013$ R$_{\odot}$.  
K-star surface brightnesses inferred from Table
3 of \citet{beuermann99} then imply that the secondary
has $M_V = 9.4 \pm 0.4$, where nearly all the uncertainty
arises from the spectral type.  Our photometry and estimated
secondary-star contribution then gives $m - M = 8.5 \pm 0.6$.
QZ Ser lies at $b = +47^{\circ}$,
and \citet{schlegel98} estimate $E(B-V) = 0.05$ in this 
location from IRAS 100-micron maps, or $A_V = 0.17$.   
Accounting for this yields a distance around 460 (+150,
$-$110) pc.  At 460 pc,
$\mu \simle  0''.03$ yr$^{-1}$ corresponds 
to $v_T \simle 65$ km s$^{-1}$, which is not unlikely.
  
{\it Orbital Parameters and Masses.} After correcting 
for the estimated blue-star contamination in
the $I$ band, the ellipsoidal light curve has $A=0.098(14)$ mag and
$\Delta A =0.015(9)$ mag.  To reproduce these numbers, we interpolated and
extrapolated from Table I of BKS, adopting plausible values of limb- and
gravity-darkening for a cool star (respectively $u=0.8$ and $\beta=0.6$, but
the results are not sensitive to these choices).  With mass ratios in the
range 0.1-0.4, we obtained the measured $A$ for inclination $i=33 \pm 4$
degrees (at such low inclinations, $A$ depends mainly on $i$).
In principle we could use the measured $\Delta A$ to constrain $q(i)$
further, but our measurement is too crude, and  
cataclysmic variables can easily produce signals at
$P_{\rm orb}$ from effects unrelated to ellipsoidal distortion.

The secondary's
velocity amplitude gives a mass function of 0.075(5) M$_{\odot}$.  
This may be distorted by illumination effects, but we see 
no evidence of variation of the line features with orbital
phase.  If we assume for a moment that $M_2 = 0.125$ M$_{\odot}$ 
and a broadly typical white dwarf mass $M_1 = 0.7$ M$_{\odot}$, the
mass function implies $i = 32^{\circ}$, essentially identical 
to the inclination derived from the ellipsoidal variations.  
We emphasize that we have not measured these masses, but 
we use them only to show that the data are consistent with 
our scenario without assuming an unusual mass for the white dwarf.  
As noted earlier, we believe that
the apparent emission line radial velocities do not indicate 
the white dwarf motion with any interesting
accuracy.

\section{Discussion}

QZ Ser is only the second short-period dwarf nova to show a K-star secondary.
In all other dwarf novae with $P_{\rm orb} \simle 2$ hr, save for a handful
of helium systems, the 
secondaries are late M dwarfs which contribute a small fraction
of the visible light.  How are we to account for this unusual object?
As with RX 2329+06 \citep{thor02}, we suggest that the 
secondary evolved significantly on the main sequence
prior to mass transfer, enhancing the core with enough helium to 
greatly affect the mass-$\te$ relation.  The small, hot secondary
we see is the remnant of a once more massive secondary.

For illustration, we computed some evolutionary models 
in the framework of the standard disrupted magnetic
braking scenario (see \citealt{bk00} and references therein). 
At the $\sim 2$ hr orbital period of QZ Ser, a donor which
started mass transfer on the zero-age main sequence
(ZAMS) would have spectral
type M4-M5 and mass $M_2 \, \sim \, 0.2 \, \msol$.  
Such a ``standard" sequence, based on the models of \citet{bk00},
is displayed in Fig.~\ref{fig4} by the solid curve. 
Also shown in Fig.~\ref{fig4} are some test calculations in which mass
transfer begins near the end of central H burning.   The
tracks shown represent
several choices of {\it initial} secondary 
masses $M_2 \, \simgr \, 1 \, \msol$, 
constant mass loss rates ${\dot M} \sim 10^{-9} - 10^{-8} \, \msolyr$,
and different levels of nuclear evolution $X_{\rm c} \simle 0.1$.
As Fig.~\ref{fig4} shows, these naturally reproduce 
the observed properties of QZ Ser and RX 2329+06.
The choice of our parameters seems reasonable and we do not require
particularly extreme assumptions on our evolutionary scenarios 
to fit these objects\footnote{With our assumptions, the
system in principle should have initially tranferred mass on a thermal 
timescale, until $M_2$
became small enough for the system to reappear as a standard
CV (see \citealt{bk00} for a discussion).}.

As already mentioned in \citet{thor02}, 
the models predict altered surface abundances. Fig. \ref{fig4}
displays the surface enrichment of $^{14}$N processed by the CNO cycle 
in the deeper layers and mixed up to the surface. The surface helium
abundance increases also, reaching mass fraction $\sim$ 0.6 at $
P_{\rm orb} \sim 2$h.  In RX 2329+06 the strengths of H$\alpha$ and 
HeI $\lambda$5876 are in the ratio 3.6:1, whereas this ratio is typically
6 and above in SU UMa stars \citep{thor02}; in the 2002 February 
spectra of QZ Ser, the ratio is 5:1, which may indicate some enhancement of
He.  As noted earlier, though, He features were nearly absent in the 
2002 January pre-outburst spectra, so the He:H line ratio is evidently
not a consistent measure of abundance in this object.  

The sodium enhancement noted earlier (if real) may provide an important
window on the secondary's nuclear processing history. 
The temperatures reached in the deepest layers of stars with masses
$\simgr \, 1.2 \, \msol$ near the end of central H burning are
high enough to allow the production of $^{23}$Na via the 
$^{22}$Ne($p, \gamma)^{23}$Na reaction.  During mass transfer,
the convective envelope proceeds inward and may reach
the Na-enriched layers.  At the 2-hour period of QZ Ser,
the bottom of the secondary's convective envelope should
reach into layers which have attained temperatures
of 1.5 -- $2 \times 10^7$ K during prior evolution.  
Based on the recent NACRE reaction rates 
\citep{angulo99}, 
this is hot enough to destroy $^{22}$Ne by proton 
capture, but it is not hot enough for the 
Na-Ne cycle to contribute significant Na enhancement \citep{weiss00}.  
A combination of deep mixing and Ne-Na cycling has been
considered as an explanation for anomalous Na abundance in 
globular cluster giants \citep{denisenko90, kraft97, weiss00}.
The nuclear reaction networks presently
implemented in our models unfortunately do not produce a 
quantitative estimate the sodium enhancement,
but work is in progress to remedy this.  
At present, we can only point to the strong Na lines
as a clue that the surface material was processed
at relatively high temperatures, mixed upward, and exposed
as mass transfer stripped away the overlying layers.

The origin of the Na enhancents in the globular cluster stars
is uncertain, and as in that case, it is possible that QZ Ser's sodium
was already present in the gas from which the system formed.
It would therefore be desirable to detect the primary products of 
CNO processing --
He and N enrichment, and C and O depletion.  Unfortunately, test
model atmospheres with $\te$ = 4300 K and $\log g = 5.5$, kindly 
computed by F. Allard (private communication)
showed essentially no observable effect when He was enhanced
to 50\% in mass fraction and the CNO abundances were altered as
expected.  More noticeable effects should appear in
models at cooler temperatures ($\te < 4000$ K),
where molecules involving C or O form. 

If our interpretation
based on nuclearly evolved donors is correct, we may expect that a 
non-negligible number of short period CVs with anomalously hot
secondaries will be discovered.  As already emphasized by \citet{beuermann98}
and \citet{bk00}, such evolved sequences 
can indeed explain a  {\it substantial fraction} of the observed CVs with late
spectral types and orbital periods $P \, \simgr \, 6$ h.  
As shown in Fig. \ref{fig4}, the most evolved
sequences which provide an explanation for such systems 
predict as well the existence of early spectral types at shorter periods.
The late spectral type systems with
$P \, \simgr \, 6$ h
represent a non-negligible fraction of systems in the sample of
\citet{beuermann98}.
\citet{bk00} were thus concerned by the fact that such evolved 
sequences may remain active in the 2-3 h period gap, since such 
extreme secondaries never become fully convective. In order to
prevent populating the period gap and predicting
too early spectral types at shorter periods,
(\citealt{bk00}, see their \S 4) suggested an 
increase of the mean mass transfer 
rate during the secular evolution of such evolved donors. However, 
the  existence of QZ Ser at $P_{\rm orb}$ = 2.0 h now supports the idea
that evolved sequences may remain active in the gap. 

Our proposed scenario would be supported if QZ Ser and 
RX 2329+06 were found to have enhanced CNO-process abundances
at their surfaces, and a more quantitative study of the abundances of 
Na, Al, and other light metals may provide this evidence despite
the low $\te$.   In the models we have computed to date, 
the secondaries in longer-period systems are significantly
less massive than in `standard' systems.  If this tendency
proves robust, accurate measures of donor masses in long-period
systems could identify systems destined to evolve into 
stars like RX 2329+06 and QZ Ser in our scenario.
Finally, if our interpretation is correct, 
one should find similar systems {\it in the period gap}. 

{\it Acknowledgments.} We gratefully acknowledge funding by the NSF 
(AST 9987334), and we thank the MDM staff for their support.  
We thank especially France Allard for computing the test model atmospheres.

\clearpage
%[This bibliography contains many references not yet called in the paper.]

\clearpage

\begin{deluxetable}{lrcc}
\tablewidth{0pt}
\tablecolumns{4}
\tablecaption{Journal of Spectroscopy}
\tablehead{
\colhead{UT Date} & 
\colhead{$N$\tablenotemark{a}} & 
\colhead{HA start}  & 
\colhead{HA end} \\
}
\startdata
2002 Jan 20 &  2 & $ -2:32$ & $ -2:20$ \\ 
2002 Jan 21 &  2 & $ -2:58$ & $ -2:45$ \\ 
2002 Feb 16 &  3 & $ -2:30$ & $ -2:00$ \\ 
2002 Feb 17 & 15 & $ -2:30$ & $ -0:13$ \\ 
2002 Feb 19 & 20 & $ -4:40$ & $ -1:27$ \\ 
2002 Feb 20 & 12 & $ -3:07$ & $ -1:21$ \\ 
2002 Feb 21 & 12 & $ -3:34$ & $ -1:20$ \\ 
2002 Feb 22 &  7 & $ -2:37$ & $ -0:57$ \\ 
2002 Jun 13 &  4 & $ +0:51$ & $ +1:18$ \\
\enddata
\tablenotetext{a}{Number of spectra.}
\end{deluxetable}

\begin{deluxetable}{lrcc}
\tablewidth{0pt}
\tablecolumns{6}
\tablecaption{Emission Features}
\tablehead{
\colhead{Feature} & 
\colhead{E.W.\tablenotemark{a}} & 
\colhead{Flux\tablenotemark{b}}  & 
\colhead{FWHM \tablenotemark{c}} \\
 & 
\colhead{(\AA )} & 
\colhead{(10$^{-15}$ erg cm$^{-2}$ s$^{1}$)} &
\colhead{(\AA)} \\
}
\startdata
         H$\gamma$ & $ 27$ & $107$ & 25 \\ 
HeI $\lambda 4471$ & $ 11$ & $ 50$ & 24 \\ 
          H$\beta$ & $ 33$ & $131$ & 25 \\ 
HeI $\lambda 4921$ & $ 10$ & $ 41$ & 36 \\ 
HeI $\lambda 5015$ & $ 13$ & $ 50$ & 45 \\ 
HeI $\lambda 5876$ & $ 11$ & $ 49$ & 25 \\ 
         H$\alpha$ & $ 36$ & $158$ & 27 \\ 
HeI $\lambda 6678$ & $  7$ & $ 29$ & 32 \\ 
HeI $\lambda 7067$ & $  9$ & $ 37$ & 37 \\ 
\enddata
\tablenotetext{a}{Emission equivalent widths are counted as positive.}
\tablenotetext{b}{Absolute line fluxes are uncertain by a factor of about
2, but relative fluxes of strong lines
are estimated accurate to $\sim 10$ per cent.} 
\tablenotetext{c}{From Gaussian fits.}
\end{deluxetable}
\clearpage

\begin{deluxetable}{lrrrrrc}
\footnotesize
\tablewidth{0pt}
\tablecaption{Fits to Radial Velocities\tablenotemark{a}}
\tablehead{
\colhead{Data set} & \colhead{$T_0$\tablenotemark{b}} & \colhead{$P$} &
\colhead{$K$} & \colhead{$\gamma$} & 
\colhead{$N$} &
\colhead{$\sigma$}  \\
\colhead{} & \colhead{} &\colhead{(d)} & \colhead{(km s$^{-1}$)} &
\colhead{(km s$^{-1}$)} & & \colhead{(km s$^{-1}$)}
}
\startdata
Absorption  & 2328.0443(3) & 0.0831612(11) &  207(5) & $-9(4)$ & 77 &  17 \\
H$\alpha$ emission & 2326.9284(12) & [0.0831612] &  69(6) & $-20(4)$ & 69 &  23 \\ 
\enddata
\tablenotetext{a}{Fits are of the form $v(t) = \gamma + K \sin[2 \pi (t - T_0)/P]$.
The number of points used is $N$ and $\sigma$ is the standard deviation from
the best fit.}
\tablenotetext{b}{Blue-to-red crossing, HJD $- 2450000$.}
\end{deluxetable}

\clearpage

\begin{figure}
% \plotfiddle{v844di_sp.ps}{7.6truein}{0}{85}{85}{-260}{-50}
\plotone{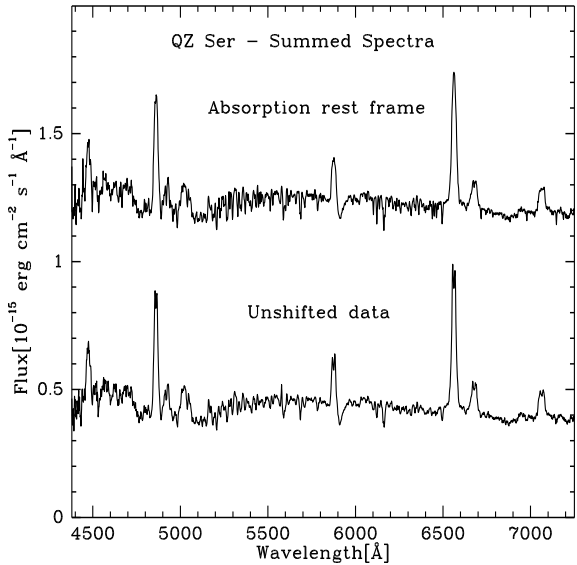}
\caption{Mean flux-calibrated spectrum.  In the upper trace,
the individual exposures have been shifted into the absorption
rest frame and the trace has been shifted upward.
}
\end{figure}

\begin{figure}
% \plotfiddle{pgrmpl.ps}{7.6truein}{0}{85}{85}{-260}{-50}
% \plotfiddle{test.ps}{7.6truein}{0}{85}{85}{-260}{-50}
% $$\BoxedEPSF{test.ps scaled 800}$$
\plotone{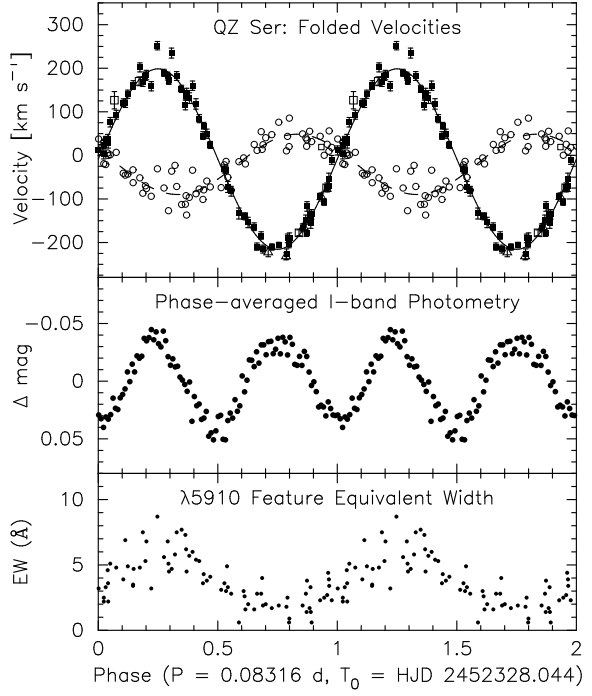}
\caption{
Data folded on the binary period.  All
points are shown twice for continuity.
{\it Upper panel:}  Radial velocities with 
best-fitting sinusoids.  Emission velocities are 
shown as round dots; absorption velocities are 
as follows: 
2002 January = open squares, 2002 February = filled squares,
and 2002 June = open triangles.
{\it Middle panel:} Phase-averaged, mean-subtracted, $I$-band 
differential magnitudes.  {\it Lower Panel:} Equivalent widths of
the diffuse $\lambda 5910$ absorption feature.
}
\end{figure}

\begin{figure}
% \plotfiddle{v844di_sp.ps}{7.6truein}{0}{85}{85}{-260}{-50}
% \plotone{figure3.ps}
\caption{
Spectra near $\lambda 5900$ rectified and 
shown as greyscale against phase.
All data are shown twice for continuity.
The emission feature to the left is HeI $\lambda$ 5876.
Note the doppler motion of the sharp absorption features,
and the diffuse absorption feature around $\lambda 5910$.
(Included as a separate jpg file in astro-ph preprint. The
white horizontal line is an artifact of the 
ps to jpg conversion.) } 
\end{figure}

\begin{figure}   
\plotone{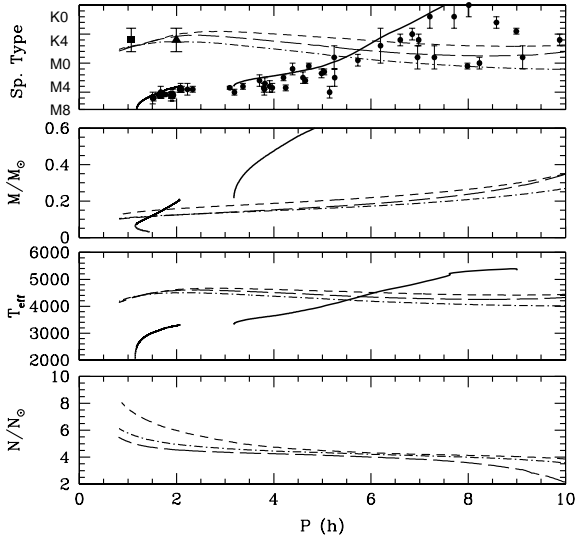}
\caption{Spectral type, mass, effective temperature and surface $^{14}N$ 
abundance (normalized to the solar abundance) of the secondary
versus orbital period. 
The solid
curve corresponds to a ``standard" sequence with an initially unevolved
donor, in the framework of the disrupted magnetic
braking model (reproducing the 2-3h period gap). 
The other curves correspond to evolutionary  sequences 
starting mass transfer
near the end of H burning, at a central H mass fraction $X_{\rm c}$. Long-dashed curve: $M_2$ = 1.2 $\msol$, ${\dot M} =
1.5 \times 10^{-9} \msolyr$, $X_{\rm c} = 4 \times 10^{-4}$.
Dash-doted curve: $M_2$ = 1.3 $\msol$, ${\dot M} =
1.5 \times 10^{-9} \msolyr$, $X_{\rm c} = 5 \times 10^{-2}$.
Short-dashed curve: $M_2$ = 1.5 $\msol$, ${\dot M} =
10^{-8} \msolyr$, $X_{\rm c} = 1.7 \times 10^{-2}$.
The locations of QZ Ser (present work, filled triangle) 
and RX 2329 \citep{thor02},
filled square) are indicated. The filled circles are observations
from \citet{beuermann98}.
}
\label{fig4}
\end{figure}

\end{document}